\documentclass[apj]{emulateapj}

\newcommand{\etal}{et~al.}
\newcommand{\fxr}{f$_X/$f$_R$}

\newcommand{\hard}{$2$--$8$~keV}
\newcommand{\soft}{$0.5$--$2$~keV}
\newcommand{\cgsflux}{erg~s$^{-1}$~cm$^{-2}$}
\newcommand{\ergss}{erg~s$^{-1}$}

\newcommand{\spitzer}{\emph{Spitzer}}
\newcommand{\chandra}{\emph{Chandra}}

\slugcomment{Accepted for publication in the Astrophysical Journal}

\shorttitle{Optically--Faint AGN with Spitzer}
\shortauthors{Rigby et al.}

\begin{document}

\title{Why Optically--Faint AGN are Optically--Faint: 
The \spitzer\ Perspective}
\author{J.~R.~Rigby\altaffilmark{1}, G.~H.~Rieke\altaffilmark{1}, 
P.~G.~P\'erez-Gonz\'alez\altaffilmark{1}, J.~L.~Donley\altaffilmark{1},
 A.~Alonso-Herrero\altaffilmark{1,2},
J.-S.~Huang\altaffilmark{3}, P.~Barmby\altaffilmark{3}, \& G.~G.~Fazio\altaffilmark{3}
}

\altaffiltext{1}{Steward Observatory, University of Arizona, 933 N. Cherry Ave., Tucson, AZ 85721}
\altaffiltext{2}{Instituto de Estructura de la Materia, Consejo Superior
de Investigaciones Cient\'{i}ficas, E-28006 Madrid, Spain}
\altaffiltext{3}{Harvard--Smithsonian Center for Astrophysics, 60 Garden Street, Cambridge, MA, 02138}


\begin{abstract}
Optically--faint X-ray sources (those with \fxr $> 10$) 
constitute about $20\%$ of X-ray sources in deep surveys, and are potentially
highly obscured and/or at high redshift.  Their faint
optical fluxes are generally  beyond the reach of spectroscopy.
For a sample of $20$ optically--faint sources in CDFS, we compile 
$0.4$--$24$~\micron\ photometry, relying heavily on \spitzer.
We estimate photometric redshifts for $17$ of these $20$ sources.
We find that these AGN are optically--faint both because they lie 
at significantly higher redshifts (median $z \sim 1.6$) 
than most X-ray--selected AGN, and because their spectra are much redder 
than standard AGN.  They have \hard\ X-ray luminosities in the Seyfert range, 
unlike the QSO--luminosities of optically--faint AGN found in shallow, 
wide--field surveys.  Their contribution to the X-ray Seyfert luminosity
function is comparable to that of $z>1$ optically--bright AGN.


\end{abstract}


\section{Introduction}
\label{sec:intro}

Deep X-ray surveys have resolved the X-ray background  
into discrete sources, verifying that it is the combined 
output of obscured and unobscured 
active galactic nuclei (AGN) (e.g., \citealt{mor03}).
The challenge now is to establish the redshift, luminosity, and column density
distributions of these AGN, and the properties of their host galaxies,
to understand AGN evolution and accretion history.
About 35\% of the X-ray detections in $1$~Ms observations
are beyond the reach of spectroscopy.    They are expected to 
be more heavily obscured and/or at higher redshift than the brighter population.  
A subset of these X-ray sources
are much dimmer in the optical, relative to their X-ray fluxes, than ordinary
AGN, and have thus been termed the ``optically--faint AGN.'' 
They are interesting in two ways.  

First, they are likely to be highly obscured: they 
lack the bright blue continua so prominent in unobscured AGN, 
and their X-ray photon indices indicate more obscuration than in Type~1 AGN.
Obscured AGN are expected to dominate the faint number counts 
and the power in the background above a few keV.

Second, these optically--faint AGN may lie at high redshift.
Pre--\chandra\ X-ray background models predicted an AGN redshift distribution 
that peaked at $z=1.3$--$1.5$ \citep{gilli99,gilli01}.   
By contrast, the redshift distribution of Chandra--selected AGN found by
spectroscopic follow-up is much lower, peaking near $z\sim0.7$ \citep{gilli}.
Models using a post--\chandra\ luminosity function (LF) can accomodate a lower--redshift
distribution \citep{ueda}, but an alternative possibility is that the 
distribution does peak at higher redshift, but significant numbers of 
high--redshift AGN have been systematically excluded from the spectroscopic surveys.

The general properties of optically--faint AGN and similar objects have been studied by 
\citet{alex_optfaint}, \citet{yan}, and \citet{koek}, but without redshift
estimation.  \citet{zheng} used optical and near-infrared photometry to 
obtain photometric redshifts for 99\% of the X-ray--selected AGN in the CDFS, 
including most of the optically-faint objects.  However, such redshifts
are extremely difficult to obtain for optically--faint sources, and
of unproven reliability.
We combine optical, near--infrared, and most importantly, 
mid--infrared (from \spitzer) photometry to obtain independent redshifts.
\spitzer\ is ideally suited to find redshifts for
these sources:  the IRAC bands (at $3.6$, $4.5$, $5.8$, and $8.0$~\micron)
are well-placed to sample the stellar emission of even very high redshift 
galaxies.  Additionally, the rest-frame near infrared (which \spitzer\ probes
for $z \ge 1$) typically offers the highest 
contrast to detect the normal stellar population against the AGN light.
Thus, the IRAC bands have the best chance of revealing stellar features that
can yield redshift determinations. 


\section{The X-ray--to--Optical Flux Ratio and Sample Selection}
\label{sec:sample_sel}

The ratio of optical R-band flux to hard X-ray 
(usually $2$--$10$~keV or $2$--$8$~keV) flux, (\fxr), can be used to classify
the emission mechanisms of X-ray sources 
(e.g. \citealt{maccacaro,comastri,barger03}).  A value $<$ 0.01 indicates 
the X-ray emission is  powered by star formation, while 0.1 $<$ \fxr ~$<$ 10 
indicates that the X-rays arise in an AGN.  
Optically--faint X-ray sources are defined to have \fxr $>$ 10,
making them poor emitters at optical wavelengths given their X-ray fluxes.
The \fxr\ ratio is defined in the observed frame, and as such is subject to 
K-corrections.

In this paper, we use the \fxr\ ratio to select optically--faint AGN.
The region sampled is the overlap between the Chandra Deep Field 
South 1~Ms Chandra observation \citep{giacconi,alex} and the GOODS
ACS optical mosaic \citep{goods_acs}.  We start by choosing objects from the
\citet{giacconi} X-ray catalog that have $2$--$10$~keV band detections
and \fxr~$>10$.   
When \citet{giacconi} list multiple R-band candidate
counterparts for an X-ray source, we require they all be optically--faint.
If no R-band counterpart is detected, we require a flux upper 
limit stringent enough to insure \fxr$>10$.  
These criteria select $48$ sources.

We then switch to the \citet{alex} CDFS X-ray catalog, since it has smaller
R--to--X-ray positional offsets than does  \citet{giacconi} catalog 
(see the appendix of \citealt{alex}.)  We do this by cross-correlating
the two X-ray catalogs, requiring hard-band detection in both catalogs 
within $1.6$\arcsec.  
This drops 9 sources from the sample:  3 of the sources have no
counterpart in the \citet{alex} catalog, even out to $5$\arcsec; and 6 are
$2$--$8$~keV non-detections (but are detected in another band) in the \citet{alex}
catalog, and thus are dropped from our sample.  
The 9 dropped sources are fainter than the optically--faint sample, with 
$2$--$8$~keV fluxes $\la 10^{-15}$~\cgsflux, as compared to the 
median $2$--$8$~keV flux for the 39 optically--faint sources of 
$3.6\times 10^{-15}$~\cgsflux.

All but 6 of the 39 optically--faint sources have 
\hard\ and \soft\ fluxes that agree within $20\%$ between both X-ray 
catalogs.\footnote{The rest have fluxes in agreement within a factor of two.}
Sources are identified by ID numbers from \citet{alex} (abbreviated AID).

%
%
%
%

To obtain a sample with high--quality SEDs, we then 
a) choose those sources that lie within the GOODS ACS field, 
which reduces the sample to $25$;
and b) require each source to have at least two photometric detections 
at wavelengths below $1$~\micron, which further reduces the sample to 
20 AGN.  
We term the resulting sample of $20$ AGN the ``complete--SED sample (CSS)''.

There are two potential sources of biases to this sample.
First, it may be somewhat brighter than the remaining optically--faint
AGN, due to the requirement for multiple--band optical detections.
Second, requiring $\lambda<1$~\micron\ detections might possibly
bias the CSS toward low redshifts compared to the full sample of
optically--faint AGN; we explore this possibility in \S~\ref{sec:koek},
by examining the redshifts of sources that would be excluded by the
$\lambda<1$~\micron\ detection requirement.

\section{\spitzer\ Observations, Photometry, and SEDs}
\label{sec:observations}

With \spitzer\ \citep{spitzer}, we obtained IRAC \citep{irac}
measurements of the CDFS with $500$~s of integration.   
The images were  
reduced by the {\it Spitzer} Science Center using the standard pipeline.
We also obtained MIPS \citep{mips}
$24$~\micron\ scan map images with a total integration time of $\sim 1200$~s
per position, nominally composed of $120$ individual sightings per source. 
These data were reduced using the instrument team data analysis tool 
\citep{gordon}, creating the image presented by \citet{x24}. 


We created a database to combine the MIPS and IRAC images with the following
optical and near--infrared imagery:
the ACS/HST \emph{bviz} images from GOODS \citep{goods_acs}; 
\emph{RIz} frames from the Las Campanas Infrared Survey \citep{marzke}; and 
the \emph{BVRI} images released by the ESO Imaging Survey \citep{arnouts};
the \emph{JK} images from GOODS \citep{goods_acs}; and the 
\emph{JK} images from the EIS Deep Infrared Survey 
at \url{eso.org/science/eis/surveys/strategy\_EIS-deep\_infrared\_deep.html}.
We also added the \chandra\ images from 
\url{astro.psu.edu/users/niel/hdf/hdf-chandra.html} \citep{alex}.

For each source, any object detected in the K band within $2$\arcsec\ of the X-ray
position was selected for photometry in all available bands.
The source selections were reviewed visually, and when necessary, were modified
so that the same source was photometered in each band.
The result is closely--sampled, deep photometry from 
$0.4$ to $8$~\micron, with additional coverage at $24$~\micron.

We now discuss the few sources that have multiple K-band components within $2$~\arcsec,
where extra care was needed to obtain accurate photometry:


AID 100:  There are four K-band sources (or components) within 
3\arcsec\ of the X-ray position, with offsets of 0.1, 1.3, 2.9, and 2.9\arcsec.
These same components are also present in the ACS z-band image.  
We quote photometry for the closest (0.1\arcsec\ offset) source.
 
AID 218:  The K-band counterpart is clearly the source located only 0.3\arcsec\ from 
the X-ray coordinates.  However, a second source 
(located 2.1\arcsec\ from the X-ray source, and 1.5\arcsec\ from the K-band counterpart)
contaminates the measured IRAC fluxes in channels 2--4.  
Therefore, we plot these fluxes as upper limits in figure~\ref{fig:complete_seds}.

AID 241:  There are two K-band components, one located 0.4\arcsec\ away,
and a fainter  source 1.6\arcsec\ from the X-ray position.  
We photometer the closer source.

AID 245:  There are two K-band components, located 0.4\arcsec\ and 1.9\arcsec\ 
from the X-ray position.
The measured IRAC fluxes are contaminated by contribution from the farther component.

AID 281:  There is a K and ACS source 0.3\arcsec\ from the X-ray coordinates; it 
appears to be extended (or double) out to  0.7\arcsec\ from the X-ray source.  We
photometer only the closer component of the extended source.


\section{Properties of the Complete SED sample}
\label{sec:css_props}

\subsection{Spectral Properties}

\begin{figure}
\figurenum{1}
\plotone{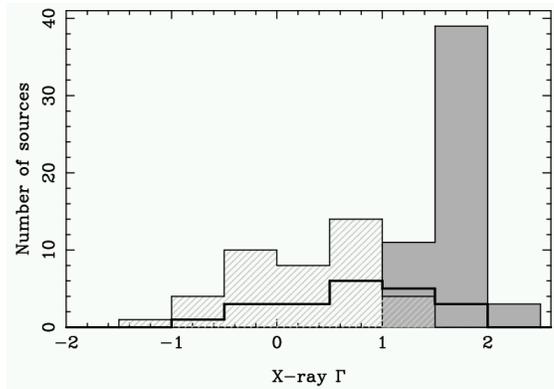}
\figcaption{Distribution of X-ray photon index $\Gamma$ (defined as
f$_{\nu} \propto \nu^{1-\Gamma}$).
Plotted are the optically--faint CSS AGN \emph{(thick line)};
X-ray--selected Type~1 AGN \emph{(shaded region)};
and X-ray--selected Type~2 AGN \emph{(cross--hatched region)}. 
Classifications are from  \citet{szokoly}.
Photon index  values are from \citet{alex}; we omit sources where $\Gamma$
is undetermined or is uncertain by more than $\pm0.5$.
}
\label{fig:gamma}
\end{figure}

Figure~\ref{fig:gamma} shows the distribution of X-ray photon index~$\Gamma$
(defined as f$_{\nu} \propto \nu^{1-\Gamma}$) for the CSS.
%
%
Unobscured AGN generally have $\Gamma \approx 2$ (and are thus flat in $\nu$f$_{\nu}$), 
whereas obscured AGN generally have $\Gamma \la 1$ 
(and thus $\nu$f$_{\nu}$ rises with increasing frequency).
The optically--faint $\Gamma$ distribution appears to be intermediate in
obscuration, with a significant number of obscured AGN.

\begin{figure}
\figurenum{2}
\includegraphics[angle=0,width=8cm]{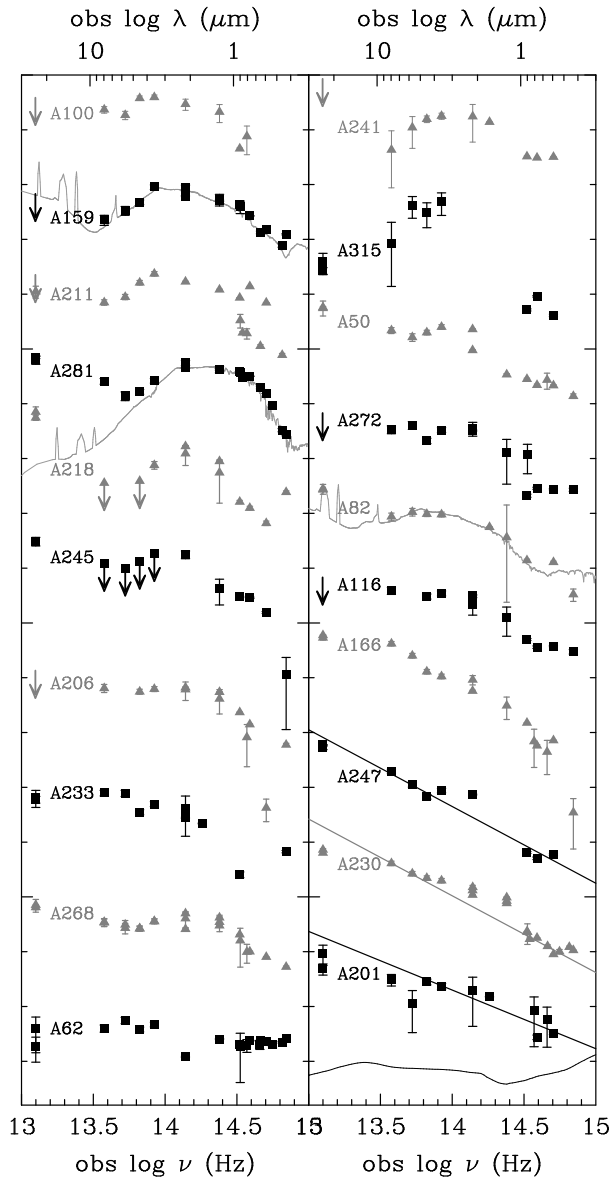}
\figcaption{The sample of complete SEDs.  Sources with strong stellar features 
are plotted first, progressing to weaker stellar features and finally to 
sources with power--law SEDs.  Wavelengths and frequencies are as observed.
For illustration, we overplot three templates from \citet{devriendt}:
M82 with an additional E(B-V)=0.2 of reddening (plotted with source AID~159); 
early-type galaxy Virgo~1003  (plotted with AID~281);
and rapidly star-forming galaxy IRAS 05189-2524 from \citet{devriendt}  (plotted with AID~82).
We also plot the median Type~1 QSO spectrum of \citet{elvis} (bottom right)
in order to illustrate how much redder are the optically--faint AGN.
}
\label{fig:complete_seds}
\end{figure}

The $0.4$--$24$~\micron\ spectral energy distributions of the CSS are plotted in 
figure~\ref{fig:complete_seds}.   
Optical through $24$~\micron\ photometry is
reported in table~\ref{tab1}; X-ray photometry is reported in table~\ref{tab2}.
All but $3$--$4$ of the SEDs show strong 
stellar features---either breaks or the characteristic broad stellar hump peaking 
near $1.6$~\micron\ rest-frame.  The remaining sources show a red,
power--law continuum, with a spectral break between $8$ and $24$~\micron.

\subsection{Redshift Techniques and Previously--Estimated Redshifts}
\label{sec:redshifts}

There are three published spectroscopic redshifts for CSS sources, 
measured by \citet{szokoly} and listed in our table~\ref{tab1}.
Of these, AID 230 and 245 have redshifts of $z=1.603$ and $z=3.064$, respectively.
Also, for AID 241 there is a published redshift of $z=0.679$ for an optical 
counterpart $1.6$\arcsec\ from the \citet{alex} position 
(which is source number 201b in \citealt{szokoly}).  However, we feel that this
is not the most likely counterpart to the X-ray source, since there is a closer,
fainter source: Szokoly source 201a, located just $0.6$~\arcsec\ from  the 
\citet{alex} position.  This closer source lacks a spectroscopic redshift.

Since only a few spectroscopic redshifts are available for the sample, one must
turn to photometric techniques.
Two groups, \citet{zheng} and COMBO-17 \citep{combo17}, have estimated photometric
redshifts for some of the CSS sources.  The former group used 
10 photometric bands from 0.3 to 2.1~\micron; the latter group used 17 passbands from 
0.35 to 0.93~\micron.  The redshift estimates from these two groups, as applicable to the
CSS, are compiled in table~\ref{tab1}.  

Adding IRAC fluxes to photometric redshift methods should dramatically improve the 
results, since IRAC samples the peak and red side of the stellar emission out to high
redshift.  Several groups are currently searching for the best way to do exactly that---the
question is not trivial, partly because of the lack of good templates over this wide
redshift range.  One solution, adopted by \citet{pabloz}, is to use empirical templates
compiled from sources in deep fields that have spectroscopic redshifts, and then
use those templates to fit photometric redshifts to other faint sources.  Their photo-z
technique is described in detail in the appendix to that paper.

Unfortunately, for galaxies that host AGN, template-fitting of any sort (whether the
high--resolution templates of standard photo-z techniques, or the low--resolution templates
used by \citep{pabloz})  can be difficult and potentially unreliable.  
First of all, each source has a differing contribution of host galaxy and AGN light.  
Secondly, some of the high-redshift, red AGN have SED shapes that are quite 
different from the templates, which are based on lower--redshift, bluer samples.

Clearly, caution is warranted when resorting to photometric redshifts, since the techniques
are unproven when applied to sources as extreme as the optically--faint AGN.  Different
techniques  of redshift determination should be tested against each other for consistency,
and the widest possible range of templates should be used.

\subsection{New Photometric Redshift Estimations}
\label{sec:ourzs}

To estimate redshifts from our 0.4--8~\micron\ data, we adopt a different,
more conservative approach than simply adopting the results of a best--template
fitting program.  Our goal is not to find the best or most
likely redshift, but rather to constrain the redshift range of each source with high
confidence.  

In our technique, we manually compared the photometry of each source 
to a range of SED templates from \citet{devriendt},
namely the early--type galaxy Virgo~1003, spiral galaxies including Virgo~1987,
the starburst galaxy M82, and the ultralumious infrared galaxy IRAS~05189-2524. 
We chose these templates to sample a wide range of star formation rate.

In a few cases, we needed to add extra reddening
(a few tenths of a magnitude in E(B-V)) to the  \citet{devriendt} templates
to match the optical fluxes, since these sources are much redder
than the \citet{devriendt} galaxies.
(For example, see AID 281 in figure~\ref{fig:complete_seds}).
We also add varying amounts of dust to make sure our redshift results
are not highly dependent on the chosen extinction.\footnote{We do not bother 
to add this absorbed UV and optical energy back to the templates
as re-emitted infrared light, since it would only contaminate the longest,
non-stellar wavelengths which are not important in stellar feature fitting.}

For each source, we compare these templates to the photometry, and find the redshift
range for which \emph{any} template provides a good match to the photometry.
For example, a source might be fit by the M82 template at $1<z<1.3$, as well as by
the M82 template with added extinction at $0.8<z<1.1$.  In this case, we would
quote the range $0.8<z<1.3$.

Our resulting redshift ranges are tabulated in table~\ref{tab1}.

Our redshift ranges agree reasonably well with the other techniques.
In 10 out of 13 cases, our redshifts agree with those of \citet{zheng}. 
We quote redshifts for four sources where \citet{zheng} did not, and
they quote redshifts for two power-law objects where we do not.
COMBO-17 lists three redshifts for CSS sources; we can estimate redshifts
for two, and agree in one case.  Three CSS sources have spectroscopic redshifts;
our photo-zs agree in two cases (AID 166 and 230), and strongly disagree 
in one case (AID 245, with z$_{spec}=3.064$, for which we securely find z=1.1--1.4).

Thus, 12 CSS sources ($57\%$) have redshifts confirmed by two different
techniques (COMBO-17 and our method, or \citet{zheng} and our method).
Also, five additional CSS sources have single--source redshift estimates at high
confidence (four from our technique, and one from \citet{szokoly}.)
Thus, 17/20 CSS sources ($85\%$) have useful redshift information.
 
We can also test the agreement between our redshift estimates and those found using
the empirical template technique of \citet{pabloz}.  For 11 of the 17 sources
with useful redshift information, the \citet{pabloz} photo-z falls within the 
quoted redshift range.  This is fairly good agreement, considering that the
empirical template technique has not been optimized to deal with AGN SEDs.

One notable problem with the photometric redshifts is the case of AID 245.
It's spectroscopic redshift of z=3.065 \citep{szokoly}, based on two 
narrow emission lines, appears to be solid.  However, we have estimated its
redshift as 1.0--1.3.
The photometry is well--measured, so why is the photometric method 
in such disagreement with the spectroscopic redshift?

We inferred a redshift of $z\sim1$ for AID~245 based on the apparent stellar
hump from $\lambda_{obs} = 1.6$--$6$~\micron, with peak at $\sim3$~\micron.
Such a spectral shape is not consistent with any stellar template at $z\sim3$
(which should peak in emission at $\lambda_{obs} \sim 6.5$~\micron.  Clearly,
the problem is not with the template--fitting, but with the photometry.  

Close examination of the images indicates that the IRAC photometry of AID~245 is 
contaminated by a neighboring source.  Thus, the photometry is probably 
a composite of two sources, which together simulate a stellar hump at $z\sim1$.
This could explain why the hump appears narrower 
(and rises more quickly from J to K-band) than stellar templates.

Thus, it appears that the redshift discrepancy for AID~245 results from
a perverse case of source blending, in that the composite SED looks stellar
(but at a fictitious redshift) rather than obviously composite.
In section \S~\ref{sec:observations}, we looked for evidence of other
blending problems, and found that cases like AID~245 are rare in the CSS sample.

\subsection{The Redshift Distribution of the Optically--Faint CSS AGN}
\label{sec:redresults}


We now consider the redshifts of the 17/20 CSS sources with useful redshift information.
When a spectroscopic redshift has been published, we use it; otherwise, we use the 
redshift ranges found using our technique in \S\ref{sec:ourzs}.

Almost all (14/17) of the CSS AGN lie at $z>1$,  
and at least $25\%$ lie at $z>2$.  
By contrast, of the $99$ reliable\footnote{``Reliable'' in 
this context means a quality flag $Q \ge 2$.}
spectroscopic redshifts available for X-ray--selected AGN in CDFS 
\citep{szokoly}, only $42\%$ lie at $z>1$;  only $17\%$ lie at $z>2$.
Thus, the optically--faint X-ray sources  lie at higher redshift than 
other X-ray--selected AGN, although their redshifts are typical of the 
optical QSO population.

Since the redshift distribution of optically--bright AGN in CDFS is strongly 
influenced by large--scale structure 
(two redshift spikes at $z=0.674$ and $0.734$), we now compare 
with several other fields.
\citet{gilli} has compiled the redshift distribution of X-ray--selected 
sources with f$_{2-10 keV} > 5\times 10^{-15}$~\cgsflux\ in CDFS, CDFN, Lockman Hole, 
Lynx field, and SSA13.  After excluding large scale structures from CDFN and CDFS,
$49\%$ of the sources lie at $z>1$, and $14\%$ lie at $z>2$.  This
distribution, too, lacks high-redshift objects compared with our results for the optically--faint AGN.

\subsection{The Luminosity Function of Optically--Faint AGN}
\label{sec:lf}

We now compute the rest--frame \hard\ luminosity function of optically--faint AGN.  
We use the standard V$_{max}$ method \citep{schmidt,hs}, 
and do not correct for incompleteness.\footnote{since we do not know the 
intrinsic V/V$_{max}$ distribution.}  
We assume $\Omega_{m}=0.27$, $\Omega_{\Lambda}=0.73$, $H_o = 72$.
Because the X-ray sensitivity varies strongly over the field, we calculate 
V$_{max}$ for each source by summing the volume contribution from each 
\chandra\ pixel in the GOODS ACS field.  
The $5\sigma$ limiting flux in each pixel was calculated by the method
of \citet{muno03}, using the CDFS exposure map \citep{alex} and 
an analytic approximation of the 
\chandra\ PSF\footnote{http://cxc.harvard.edu/chandra-users/0195.html}.

\begin{figure}
\figurenum{3}
\plotone{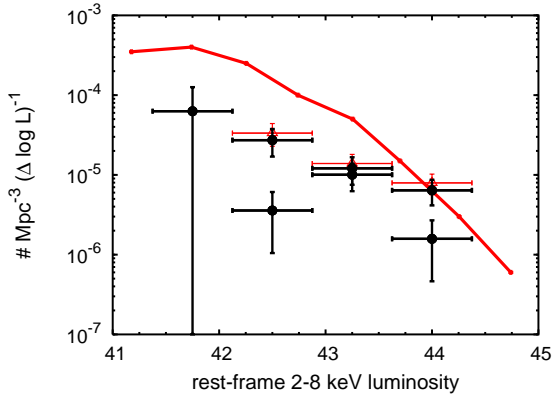}
\figcaption{The rest--frame \hard\  AGN luminosity function.  
Filled circles show the optically--faint AGN LF, computed using the 17 CSS
AGN with constrained redshifts.  We compute this LF twice, once using the lowest 
permitted redshift for each source, and again using the highest redshift.
The LF has been corrected for known AGN lacking redshifts (see \S~\ref{sec:lf}); 
we do not otherwise correct for incompleteness.  
For comparison, we plot the $0.1<z<1$ LF \emph{(red solid line)} 
for AGN with spectroscopic redshifts \citep{steffen}.
We also compute the LF for those $1<z<4$ AGN in the GOODS field with 
\citet{szokoly} redshifts \emph{(red triangles)}.
}
\label{fig:lf}
\end{figure}

Figure~\ref{fig:lf} plots the resulting LF, based on
the redshift constraints given in table~\ref{tab1} (using the
spectroscopic redshift if available, and otherwise our redshift range found 
in \S\ref{sec:ourzs}.)
Within the GOODS ACS field, there are $27$ optically--faint sources with 
high-confidence hard-band detections (detected by both \citealt{alex} and 
\citealt{giacconi}); of these, we found redshift information for $17$.
Therefore, we have multiplied the optically--faint AGN LF found for the $17$
sources by $27/17$ (which assumes the redshift distributions are similar.)

For comparison, we plot the spectroscopically--determined rest--frame \hard\ LF 
for X-ray--selected AGN from $0.1<z<1.0$ \citep{steffen}.
Two general conclusions are apparent. 
First, the optically--faint AGN are not terribly luminous---most have 
$\log L_x \sim 42.5$--$44$ erg~s$^{-1}$.  Thus, we sample
lower luminosities than the optically--faint QSOs discovered in 
shallow, wide-field surveys, \citep{fiore,mignoli,brusa}
as expected given the survey sensitivities.
Second, the number density of the optically--faint 
AGN in CDFS is comparable to that of $1<z<4$ AGN that have been identified 
by optical spectroscopy.  Therefore, optically--faint AGN boost the 
high--redshift tail of the X-ray--selected AGN redshift distribution, but 
only by a factor of $\sim2$.

Thus, we find that the optically--faint sources have moderate redshifts
($1<z<3$) and Seyfert luminosities (log L$_{2-8 keV} < 44$~\ergss), rather
than higher redshifts and QSO luminosities.  As such, they are not
expected to contribute strongly to the X-ray background, as compared
to $z<1$ Seyferts and $z>2$ QSOs (see for example figures 16 and 17 of
\citet{ueda}).
Several techniques are presently being tested to identify additional
high--obscuration AGN (for example, see \citet{almudena05} and
\citet{donley05}.  When results from such studies are available, it will
be valuable to determine the overlap between selection methods, calculate
the discovered sources' contribution to the X-ray background (as well as
the contribution from the optically--faint AGN), and re-evaluate what
fraction of the obscured AGN have been identified.

\section{Other Optically--Faint Sources}
\label{sec:koek}
Some sources, especially at high redshift, may be too faint to be
identified by our \fxr\ flux criterion.  Further, absorption from the 
Lyman~$\alpha$ forest will suppress the R-band flux for $z\ga 4$.

Therefore, we examine the most extreme optically faint cases:  
X-ray detections that are undetected even in extremely deep
z'~$>28$ imaging with HST.
\citet{koek} have identified seven such sources in the CDFS/GOODS field,
and suggested they may lie at $z>6$ or be very dusty.
To test the nature of these AGN, we have combined
the \spitzer\ bands with the GOODS photometry (\citet{koek}).
We find several source categories: 1.) Source K4 is an X-ray detection, 
but an otherwise blank field---lacking an
optical, near--infrared, or \spitzer\ counterpart;  
2.) Sources K1 and K6 have power--law SEDs; and 
3.) Sources K2, K3, and K7 have stellar-featured SEDs, with nominal 
photometric redshifts of $z\sim3$, $z\sim2$, and $2<z<6$ respectively.
Source K5 is intermediate between these last two categories.



A number of optically--faint sources outside the CSS have 
steep power--law SEDs:  AID 4, 42, 45, 64, 79, 92, and 283.  
They may be related to the optically--faint, power-law
members of the  flat spectrum 1~Jy (at 5~GHz) sample (\citet{stickel}).
The faint 1 Jy power-law sources have spectral indices  $\alpha$ $\ge -2.5$ 
(with $\alpha$ defined as f$_{\nu} \propto \nu^{\alpha}$).
The spectral indices of our objects are listed in table~\ref{tab1};
they are generally in agreement with the limit derived by \citet{stickel} for
the radio sample. The steep power law sources comprise only
about 2\% of the flat spectrum 1~Jy sample; their incidence in the CDFS appears to
be as high or higher. 

\section{Conclusions}

We report reliable (e.g. confirmed by independent methods) photometric redshifts for a
representative sample of optically--faint X-ray sources in the CDFS.  
We find that they have higher redshifts ($z>1$) than most X-ray--selected AGN,
but only $\sim30\%$ lie at $z>2$.  Thus, they populate the redshifts where
optical QSOs are most numerous.
Their $0.4$--$24$~\micron\ SEDs are intrinsically redder than typical AGN, and notably
lack bright blue continuua.  Their X-ray spectra indicate  significant absorption.  
Their X-ray luminosities are modest 
($\log L_x \sim 42.5$--$44$ erg~s$^{-1}$), and they boost the high--redshift tail of
the X-ray--selected AGN distribution by a factor of 2.

\acknowledgements
This work is based in part on observations made with \emph{Spitzer}, 
which is operated by the Jet Propulsion Laboratory, 
California Institute of Technology under NASA contract 1407. Support for 
this work was provided by NASA through Contract Number 960785 issued by 
JPL/Caltech.



\clearpage
\begin{deluxetable}{lllllllllll}
\tabletypesize{\tiny}
\tablecolumns{11}
\tablecaption{Multi-band Photometry and Redshifts for the Optically--Faint AGN Sample.\label{tab1}}
\tablehead{
\colhead{AID/XID}  &  \colhead{optical} & \colhead{Ks} &
\colhead{$3.6$~\micron} & \colhead{$4.5$~\micron} & \colhead{$5.7$~\micron} & \colhead{$8.0$~\micron} & 
\colhead{$24$~\micron} & \colhead{$\alpha$}  & \colhead{Lit. redshift}   &  \colhead{Our redshift}\\
}
\startdata  
50/227  &  R$=0.24\pm0.08$   &$4.7\pm2$     &  $12 \pm 1$     & $12 \pm 1$     & $12.4 \pm 2$   & $23.0 \pm 3$  & $180 \pm 55$     & -2.3  &   2.18 (1.78-2.54)$^1$    &   2.8--3.5\\
62/64   &  R$=0.76\pm0.07$   &$1.39\pm0.03$ &  $8.9 \pm 0.9$  & $9 \pm 1$      & $17 \pm 2$     & $17 \pm 2.4$  & $49 \pm 30$      & -0.88 &   $1.27\pm0.2^2$          &   \nodata\\
82/58   &  z$=0.26\pm0.01$   &\nodata       &  $7.2 \pm 0.7$  & $9 \pm 1$      & $1.3 \pm 2$      & $15 \pm 2$    & $141 \pm 28$   &  -2.0 &   0.92 (0.58-1.22)$^1$    &   1.8--3.8\\  
100/82  &  i$=0.65\pm0.3$    &$6.7\pm3$     &  $15 \pm 1.4$   & $18 \pm 1.7$   & $11 \pm 2$     & $20 \pm 3.0$  & $<98$	           &       &   1.89 (1.69-2.05)$^1$    &   1.1--1.8\\
116/205 &  i$=0.15\pm0.01$   &$3.1\pm0.6$   &  $6.6 \pm 0.7$  & $7.3 \pm 0.8$  & \nodata        & $16 \pm 2$    & $<86$	           &  -2.3 &   1.56 (1.31-2.3) $^1$    &   1.3--1.5\\
159/48  &  R$=0.46\pm0.03$   &$8.5\pm1.5$   &  $17 \pm 2$     & $11.6 \pm 1$   & $10.1 \pm 2$   & $9.6 \pm 2$   & $<86$	           &       &   1.26 (1.03-1.49)$^1$    &   0.7--1.1\\
166/45  &  R$=0.11\pm0.07$   &$6.0\pm1.4$   &  $14 \pm 1$     & $21.8 \pm 2$   & $53.9 \pm 5$   & $125 \pm 12$  & $480 \pm 45$     & -2.6  &   2.29 (2.14-2.60)$^3$    &   1.0--2.5\\
201/515 &  R$=0.13\pm0.09$   &$1.4\pm1.1$   &  $2.8 \pm 0.4$  & $4.2 \pm 0.5$  & $2.2 \pm 1.5$  & $8.2 \pm 2$   & $75 \pm 30$      & -2.1  &   2.19 (2.15-2.45)$^1$    &   1.3--4.8\\
206/265 &  i$=0.29\pm0.01$   &$3.6\pm0.2$   &  $5.9 \pm 0.6$  & $6.7 \pm 0.7$  & \nodata        & $14 \pm 2$  & $<86$	           &       &   1.16 (1.02-1.32)$^1$    &   1.0--1.4\\
211/35  &  R$=0.39\pm0.02$   &$19\pm0.4$    &  $44 \pm 4$     & $39 \pm 4$     & $27 \pm 3$     & $30 \pm 3$    & $140 \pm 35$     &       &   $1.14\pm0.14^2$         &   0.9--1.4\\
218/148 &  i$=0.064\pm0.005$ &$2.1\pm0.3$   &  $1.8 \pm 0.3$  & $<1.2$         & \nodata        & $<1.9$        & $112 \pm 27$     &       &   1.74 (1.50-2.02)$^1$    &   \nodata\\  
230/31  &  R$=0.55\pm0.03$   &$18.8\pm3$    &  $47 \pm 4$     & $66. \pm 6$    & $100 \pm 9$    & $217 \pm 20$  & $1008 \pm 62$    &       &   1.603$^4$; $1.1\pm0.1$$^2$ &  1.5--2.0\\
233/79  &  z$=0.097\pm0.01$  &$3.1\pm0.6$   &  $7.4 \pm 0.7$  & $6.7 \pm 0.7$  & $19 \pm 2$     & $26 \pm 3$    & $65 \pm 22$      & -1.8  &   1.91 (1.77-1.97)$^1$    &   1.0--2.5\\
241/201 &  i$=0.098\pm0.007$ &$1.6\pm1$     &  $2.7 \pm 0.3$  & $3.0 \pm 0.4$  & $2.6 \pm 1.5$  & $1.4 \pm 1.4$ & $<86$	           & -0.90 &   0.679$^4$\tablenotemark{a}; 0.14, 1.0$^2$ &   1.5--2.2\\
245/27  &  i$=0.47\pm0.02$   &$8.0\pm1.3$   &  $<14$          & $<12.6$        & $<12$          & $<20$         & $155 \pm 28$     & -2.0  &   3.064$^4$               &   1.0--1.4\\
247/25  &  i$=0.21\pm0.01$   &$8.4\pm0.1$   &  $16.5 \pm 2$   & $16 \pm 1.5$   & $34 \pm 4$     & $82 \pm 8$    & $660 \pm 50$     & -2.4  &   2.26 (1.89-2.58)$^1$    &   1.7--4.7\\
268/147 &  i$=0.41\pm0.01$   &$4.4\pm1$     &  $6.8 \pm 0.7$  & $6.3 \pm 0.7$  & $8.1 \pm 2$    & $15 \pm 2$    & $90 \pm 23$      & -2.0  &   0.99 (0.79-1.21)$^1$    &   0.8--1.1\\  
272/146 &  i$=0.12\pm0.01$   &$3.9\pm0.2$   &  $6.2 \pm 0.6$  & $5.0 \pm 0.6$  & $12 \pm 2$     & $14 \pm 2$    & $<86$	           & -2.5  &   2.67 (2.47-2.85)$^1$    &   2.4--3.4\\
281/159 &  R$=1.7\pm0.08$    &$14.6\pm1.5$  &  $12 \pm 1$     & $9.9 \pm 1$    & $11 \pm 2$     & $27 \pm 3$    & $200 \pm 36$     &       &   3.30 (3.04-3.62)$^1$    &   0.2--0.6\\
315/506 &  i$=1.0\pm0.1$     &\nodata       &  $260 \pm 115$  & $210 \pm 100$  & $360 \pm 150$  & $100 \pm 100$ & $140 \pm 60$     & -2.9  &   3.69 (3.12-4.19)$^1$    &   \nodata\\
\enddata
\tablecomments{Column 1:  AID/XID are the source identification numbers in the \citet{alex} and \citet{giacconi} 
catalogs, respectively.  Column 2 is optical flux density: we quote R-band when available, then i, then z.
Columns 2--8:  optical, Ks, IRAC, and MIPS flux densities are all listed in $\mu$Jy.  
Column 9:  the $0.4$--$8$~\micron\ spectral index $\alpha$ is defined in the text.
Column 10: ``Lit. Redshift'' is the source redshift found in the literature.  
Column 11:  ``Our Redshift'' is the redshift found using our techniques, as described in the text.
REFERENCES--  
[1] Photometric redshift using BPZ and HyperZ, from \citet{zheng};  
[2] Photometric redshift from Combo-17 \citep{combo17};  
[3] Redshift from single--line spectrum \citep{szokoly} and HyperZ, from \citet{zheng};  
[4] Spectroscopic redshift from \citet{szokoly}.\\  
\tablenotetext{a}{Reshift is for optical source 201b listed by \citet{szokoly}.  We suspect the true optical counterpart is Szokoly 201a, which lacks a spec. redshift.}
}

\end{deluxetable}

\begin{deluxetable}{lllll}
\tablecolumns{5}
\tablecaption{X-ray Photometry for the Optically--Faint AGN Sample.\label{tab2}}
\tablehead{
\colhead{AID/XID}  &  \colhead{offset} & \colhead{0.5--2~keV flux}  &  \colhead{2--8 keV flux}  &
\colhead{X-ray $\Gamma$}\\ 
}
\startdata   
50/227   &  $0.4$& $1.2 \pm 0.4$ &  $36.2 \pm 5.2$   &     $-0.43^{+0.31}_{-0.32}$\\
62/64    &  $0.2$& $21.7 \pm 1.2$ &  $55.2 \pm 4.7$  &     $1.36 \pm 0.09$\\        
82/58    &  $0.2$& $7.1 \pm 0.7$ &  $19.6 \pm 3.1$   &     $1.30^{+0.18}_{-0.16}$\\ 
100/82   &  $0.4$& $2.4 \pm 0.4$ &  $15.1 \pm 2.8$   &     $0.71^{+0.24}_{-0.22}$\\ 
116/205  &  $0.3$& $1.5 \pm 0.5$ &  $14.4 \pm 4.1$   &     $0.41^{+0.41}_{-0.37}$ \\
159/48   &  $0.9$& $9.1 \pm 0.9$ &  $48.6 \pm 5.4$   &     $0.83 \pm 0.13$       \\ 
166/45   &  $0.4$& $10.9 \pm 0.9$ &  $49.4 \pm 4.9$  &     $0.94^{+0.12}_{-0.11}$ \\
201/515  &  $0.3$&  $0.9 \pm 0.3$ &  $15.1 \pm 3.0$  &     $0.02^{+0.33}_{-0.31}$ \\
206/265  &  $0.5$& $2.2 \pm 0.5$ &  $40.8 \pm 5.8$   &     $-0.08 \pm 0.25$ \\      
211/35   &  $0.9$& $40.8 \pm 6.4$ &  $142.0 \pm 35$&       $1.13^{+0.28}_{-0.24}$ \\
218/148  &  $0.5$& $3.9 \pm 0.5$ &  $28.5 \pm 3.8$   &     $0.61^{+0.18}_{-0.17}$ \\
230/31   &  $0.3$& $59.7 \pm 1.9$ &  $87.8 \pm 5.2$  &     $1.75 \pm 0.06$ \\       
233/79   &  $0.2$& $8.8 \pm 0.9$ &  $15.5 \pm 2.8$   &     $1.62^{+0.19}_{-0.18}$ \\
241/201  &  $0.3$& $4.6 \pm 0.6$ &  $21.4 \pm 3.4$   &     $0.93^{+0.19}_{-0.18}$\\ 
245/27   &  $0.3$& $7.5 \pm 0.7$ &  $70.6 \pm 5.6$   &     $0.42 \pm 0.11$ \\       
247/25   &  $1.0$& $5.4 \pm 0.7$ &  $93.4 \pm 7.9$   &     $-0.02 \pm 0.14$\\       
268/147  &  $0.1$& $1.9 \pm 0.4$ &  $72.6 \pm 7.0$   &     $-0.61 \pm 0.22$\\       
272/146  &  $1.1$& $4.6 \pm 0.6$ &  $25.5 \pm 4.0$   &     $0.81^{+0.19}_{-0.18}$ \\
281/159  &  $0.7$& $24.0 \pm 1.3$ &  $75.3 \pm 5.7$  &     $1.21 \pm 0.08$ \\       
315/506  &  $0.7$& $6.9 \pm 0.9$ &  $9.3 \pm 3.7$    &     $1.81^{+0.45}_{-0.33}$\\   
\enddata
\tablecomments{Column 1:  AID/XID are the source identification numbers in the \citet{alex} 
and \citet{giacconi} catalogs, respectively.  
Column 2:  ``Offset'' is the offset (in \arcsec) between the \citet{alex} and 
\citet{giacconi} X-ray coordinates.
Column 3--4:  X-ray fluxes are quoted from \citet{alex}, and have units of 
$10^{-16}$~erg~s$^{-1}$~cm$^{-2}$. 
Column 5:  The X-ray photon index $\Gamma$ is defined in the text.    
}
\end{deluxetable}



\end{document}